\documentclass{aa}  

\usepackage{graphicx}
\usepackage{multicol}
\usepackage{txfonts}
\usepackage{hyperref}

\usepackage[switch]{lineno} 

\newcommand\ed{\tilde{\delta}} 
\newcommand\edl{\tilde{\delta}_{\ell}}

\usepackage{xcolor}

\begin{document} 


\title{Mass scaling relations for dark halos \\from an analytic universal outer density profile}
   \author{Giorgos Korkidis
         \inst{1}\fnmsep\inst{2}\fnmsep
          \and
          Vasiliki Pavlidou 
  \inst{1}\fnmsep\inst{2}
          }
    
    \authorrunning{ Korkidis  \&  Pavlidou}
    \titlerunning{Mass scaling relations from an analytic universal outer cluster density profile}

   \institute{Department of Physics and Institute for Theoretical and Computational Physics, University of Crete, GR-70013 Heraklio, Greece
   email:{\tt{gkorkidis@physics.uoc.gr;pavlidou@physics.uoc.gr}}
         \and
             Institute of Astrophysics, Foundation for Research and Technology – Hellas, Vassilika Vouton, GR-70013 Heraklio, Greece
             }

   \date{}

 
  \abstract
   {The average matter density within the turnaround scale, which demarcates where galaxies shift from clustering around a structure to joining the expansion of the Universe, is an important cosmological probe.  However, a measurement of the mass enclosed by the turnaround radius is difficult. Analyses of the turnaround scale in simulated galaxy clusters place the turnaround radius at about three times the virial radius in a \(\Lambda CDM\) universe and at a (present-day) density contrast with the background matter density of the Universe of \(\delta \sim 11\). Assessing the mass at such extended distances from a cluster's center is a challenge for current mass measurement techniques. Consequently, there is a need to develop and validate new mass-scaling relations, to connect observable masses at cluster interiors with masses at greater distances.}
   {Our research aims to establish an analytical framework for the most probable mass profile of galaxy clusters, leading to novel mass scaling relations, allowing us to estimate masses at larger scales. We derive such analytical mass profiles and compare them with those from cosmological simulations.}
   {We used excursion set theory, which provides a statistical framework for the density and local environment of dark matter halos, and complement it with the spherical collapse model to follow the non-linear growth of these halos.}
   {The profile we developed analytically showed good agreement (better than 30\%, and dependent on halo mass) with the mass profiles of simulated galaxy clusters. Mass scaling relations were obtained from the analytical profile with offset better than 15\% from the simulated ones. This level of precision  highlights the potential of our model for probing structure formation dynamics at the outskirts of galaxy clusters.}
   {}


   \maketitle
%

\section{Introduction} \label{section 1}
Galaxy clusters, the largest gravitationally bound structures in the universe, have long been recognized as valuable cosmological laboratories. During the past three decades, significant advances have been made in understanding their composition, dynamics, and integral role within the cosmic web. The well-studied interiors of these clusters, particularly their relaxed regions, have laid the foundations for our concordance model of hierarchical structure formation. This progress has now extended beyond the traditionally studied virialization regime to include the kinetically driven splashback regions \citep{Splashback1,Splashback5, Splashback2, Splashback3, Splashback4}. However, the potential of the outer regions of galaxy clusters, extending outside the splashback radius, into the infalling region and beyond, remains largely untapped. These outer reaches hold substantial promise for deepening our understanding of structure formation and cosmology.

Recently, the turnaround scale has received significant attention in cosmological studies \citep{Turnaround_graveyard1, Turnaround_graveyard2, Turnaround_graveyard3, Turnaround_graveyard4,Turnaround_graveyard5}. This scale represents the point at which galaxies transition from infall towards the central cluster to expansion with the background Universe. In previous work \citep{PavlidouEtal2020}, we identified the turnaround scale as a novel cosmological probe. We showed that the present average matter density on this scale (the turnaround density $\rho_{ta}$) probes the overall matter content of the universe, and that its evolution with time is influenced by the existence of a cosmological constant. The turnaround scale has thus the potential to provide new constraints on cosmological parameters, complementary to the ones obtained from the Cosmic Microwave Background (CMB)(e.g., \citealp{CMB-WMAP1}), Baryon Acoustic Oscillations (BAO)(e.g., \citealp{BAO1}), and Type Ia Supernovae (e.g., \citealp{SNIa}).

Our subsequent studies \citep{KorkidisEtAl, Korkidis_etal2023} confirmed the utility of $\rho_{ta}$ for testing cosmological models using N-body cosmological simulations. However, measuring the turnaround density in actual observations using current astronomical surveys presents significant challenges. The radius where galaxies join the Hubble flow has thus far been measured for only a few nearby superclusters \citep{ TA_measurements2,TA_measurements3, TA_measurements1}. Accurately measuring the total mass of galaxies at these scales presents an even more formidable challenge. Several issues complicate such observations on a cluster-by-cluster basis, making the associated errors unsuitable for precise cosmological analysis: (a) Accurate measurement of the mass at the turnaround scale would necessitate extensive spectroscopic and gravitational lensing surveys, mapping the galaxy distribution on very large scales around clusters. (b) In the context of the cold dark matter (CDM) model of structure formation, it is well-acknowledged that most of the mass surrounding galaxies is dark and, for such small objects as galactic halos, largely inaccessible; (c) Foreground and background galaxy contamination at these scales is non-negligible and can skew observations.

In this context, developing scaling relations that would allow us to infer the mass at turnaround scales from well-established observable masses becomes paramount. Current observational techniques, such as X-ray, Sunyaev-Zel'dovich, and weak-lensing surveys \citep{Chandra_masses, eRossita, SZ_clusters1, SZ_Clusters2, SZ_Clusters3, SZ_Clusters4, SZ_Clusters5, SZ_Clusters6}, in conjunction with machine learning methods (see, e.g., \citealp{ML_Masses1, ML_Masses2, ML_Masses3, ML_Masses4, ML_Masses5}), routinely identify clusters and measure fixed-overdensity masses (masses on a scale where the average density of the cluster is a set multiple of the the mean matter density in the universe). These overdensity masses scale well with each other, leading to the plausible hypothesis that a straightforward scaling relation might exist among all overdensity masses. The turnaround mass is also  a fixed-overdensity mass (for instance, in a $\Lambda$CDM cosmology at \(z=0\), the turnaround mass corresponds to an overdensity of \(\delta=11\), see for example \citealp{KorkidisEtAl}). Identifying the origin of these scalings could thus provide a straightforward path towards establishing scalings for the turnaround mass as well. 

One path to developing a predictive model for these scaling relations may pass through the the mass profile of cluster halos. Much of the effort in this field focuses on the cluster interior, resulting in the widespread use of common profiles like NFW or Einasto \citep{NFW,Einasto}. These profiles were developed to describe the innermost parts of such structures through extensive studies in cosmological simulations. More recently, \citet{Diemer_profile} has proposed a model that fits robustly the entire stacked density profile of simulated clusters, both interiors and exteriors.
In practice, such comprehensive profiles could be leveraged to construct scaling relations. 

In this study, we do exactly this, making use of 
a model for the mass profile of cluster exteriors derived analytically from first principles. 
The advantages of using an analytic model for the profile are twofold. First, the derivation process elucidates the physical processes that shape the profile. Second, the profile parameters are directly and causally relatable to cosmological and structure-formation parameters, without the need for expensive runs of large suits of cosmological simulations.

To derive an analytic relation for the (most probable) outer density profile of galaxy clusters, we make use of the model for the density distribution around dark matter halos proposed by \cite{PF05}. Our aim is to arrive to an explicit relation between the overdensity mass in the inner, relaxed portions of cluster-sized halos and the overdensity mass in their outer regions, including the turnaround scale. 

This paper is organized as follows: In section \ref{section 2}, we develop an analytic formulation for the most-probable outer mass profile of galaxy clusters, and we illustrate how mass scaling relations can be directly derived from this formulation. In section \ref{section 3}, we detail the cosmological simulations employed in our study to test our analytic profile, and describe the methodology used to evaluate the performance of our analytical models. In section \ref{section 4} we compare our derived profile with those obtained from simulations and test the validity of the associated mass scaling relations. Finally, in section \ref{section 5}, we summarize our findings.

\section{An analytic closed-form profile for cluster exteriors} \label{section 2}

To build a model of the density distribution around cosmic structures of varying masses,  \cite{PF05} used excursion set theory to derive a joint distribution of structures with respect to mass and surrounding overdensity, 
thus providing a general, two-parameter statistical description. This double distribution extends the Press-Schechter formalism by introducing a clustering scale parameter, $\rm \beta$, which quantitatively defines the "environment" around a structure of mass $m$ to be a scale that includes mass $\beta m$ (including the structure itself, so $\beta> 1$).  By providing the statistics of the overdensity field as a function of enclosed mass, the double distribution offers a path to deriving the most probable density profile at large distances away from cosmic structures - in other words, to understanding how interior cluster profiles merge into the background density of the expanding Universe. 

Models of the outer average density profile of clusters from first principles were pioneered by \citet{Barkana,PradaEtal, Betancort-RijoEtal06, TavioEtal}. Our methodology, while not fundamentally divergent from theirs, differs in four key choices.  Firstly, in contrast to their use of a 'Lagrangian variable $q$'—essentially the enclosed mass in units of radius— as an independent variable, we use the enclosed mass itself. This choice not only allows for a more intuitive understanding of our derivations, but, most importantly,  straightforwardly leads to usable mass scaling relations. Secondly, we employ a much simpler, while still accurate, approximation for the transformation of linear-theory overdensities to spherical-collapse ones, which was proposed by \citet{PF05}. This choice simplifies the calculations and results in a more transparent closed-form expression. Thirdly, our focus is primarily on the largest structures. This specificity allows us to omit, with minimal error, corrections for structures engulfed by larger ones, given the immense size and rarity of our targeted structures. Finally, we approximate the variance of the density field, $S(m)$, with a power law of the smoothing mass scale, $m$. This approximation holds when concentrating on a limited range of masses, as we do in our study, and it is critical in demonstrating the (quasi-)universality of our derived profile.
The cumulative effect of these modifications is a significant simplification of the mathematical framework without a significant compromise in predictive power. 

Building upon the work by \cite{PF05}, we extend their analysis by deriving the most probable density profile as a function of collapsed mass from the double distribution. In what follows, we summarize the mathematical formalism and the assumptions employed in this derivation. 

\subsection{The double distribution}
 In \citet{PF05}, a joint distribution is constructed to describe the frequency, within a given volume, of collapsed structures in specific intervals of mass and local overdensity. The overdensity is defined as the density contrast with respect to the average matter-density of the background Universe, calculated on a smoothing scale enclosing mass \(\beta m\) (including the central structure itself). It is labelled by the corresponding linearly extrapolated to the present time overdensity, denoted by \(\ed_{\ell}\). Mathematically, then, the distribution provides the comoving number density of collapsed, relaxed halos with mass between \([m, m + dm]\), residing within local overdensity in the range \([\ed_{\ell}, \ed_{\ell} + d\ed_{\ell}]\). The linearly extrapolated overdensity field $\ed_{\ell}$ corresponds to the overdensity field that would result if all structures continued to grow according to linear theory until the present time. The mathematical representation for the double distribution is the following:

\begin{align} \label{dd}
    \frac{dn}{dmd\ed_{\ell}}(m,\ed_{\ell},\beta,a) &=
\frac{\rho_{\rm m,0}}{m} \,\,
\frac{
\exp\left[-\frac{\ed_{\ell}^2}{2S(\beta m)}\right]
-\exp \left[\frac{(\ed_{\ell}-2\ed_{\rm 0,c}(a))^2}{2S(\beta m)}\right]}
{\sqrt{2\pi S(\beta m)}}
\\ \nonumber 
& \times 
\frac{\left[\ed_{\rm 0,c}(a)-\ed_{\ell}\right]
\exp\left[-\frac{\left(\ed_{\rm 0,c}(a)-\ed_{\ell}\right)^2}{2\left[S(m)-S(\beta m)\right]}\right]}
{\sqrt{2\pi}\left[S(m)-S(\beta m)\right]^{3/2}}
\left|\frac{dS}{dm}\right|_m \,.
\end{align} 
In this expression, $\ed_{\rm 0,c}(a)$ is the linearly extrapolated overdensity of a structure collapsing at redshift $z$ (scale factor $a=1/(1+z)$), extrapolated to today; $\ed_\ell$ is the linearly extrapolated overdensity of a sphere enclosing the central collapsed structure, extrapolated to the present time; and $S(m)$ is the variance of the density field when smoothed on a scale that encloses mass $m$. One implicit assumption made in Eq.~\ref{dd} is that for steps in mass \(\rm \Delta m\), the density variance \(\rm \Delta S(m)\) (and hence, the density contrasts \(\Delta \ed_{\ell} (m))\) are uncorrelated (see Appendix ~\ref{sigma}, for a more involved discussion).

The use of linearly extrapolating overdensities in the derivation of Eq.~(\ref{dd}) is necessary for the treatment of halo statistics to remain analytically tractable.  
However, before Eq.~(\ref{dd}) can become predictive for structures residing in the real, non-linear density field, we have to relate these linearly extrapolated overdensities to their non-linear counterparts. To this end, we use a relation introduced in \citet{PF05} which connects linearly extrapolated overdensities $\ed$ to overdensities $\delta$ determined using the spherical collapse model:
\begin{equation}\label{magic}
\ed_a \approx \ed_{\rm c}\left[1-(1+\delta_a)^{-1/\ed_{\rm c}}\right]\,,
\end{equation}
In Eq.~(\ref{magic}), $\ed_a $ is an overdensity linearly extrapolated to time $a$, $\delta_a$ is the real overdensity at that same time $a$, and $\ed_{\rm c}$ is the overdensity of a collapsing structure linearly extrapolated to the time of collapse. The central assumption of the spherical collapse model in that the matter surrounding each overdensity is distributed isotropically and follows radial motion during its collapse. This relation is universal for all structures and all times of collapse. Eq.~(\ref{magic}) has the correct asymptotic behavior both at low and high values of $\delta_a$, and is accurate to few percent throughout its domain.

\subsection{From the double distribution to the most probable profile}

Another way to state the content of Eq.(\ref{dd}) is that, for a given central collapsed structure of mass $m$ at a cosmic epoch $a$, the double distribution describes the probability distribution of over-densities $\ed_\ell$ around structures of that mass $m$, in spheres of increasing mass $\beta m$. We can then derive a profile (a function of $\beta m$) of maximum-probability (linearly extrapolated) overdensity - the maximum probability profile. Mathematically, this profile is simply derived by differentiating the double distribution with respect to overdensity, and setting that derivative to zero. 

In what follows, we describe a series of simplifications, both to the double distribution itself, and to the profile derivation process, that result in a simple, intuitive, closed-form profile, without significant loss of accuracy. 

{\bf Simplification 1: dropping the structure-in-structure correction.}
In Eq.~(\ref{dd}), the factor:
\[
\frac{
\exp\left[-\frac{\ed_{\ell}^2}{2S(\beta m)}\right]
-\exp \left[\frac{(\ed_{\ell}-2\ed_{\rm 0,c}(a))^2}{2S(\beta m)}\right]}
{\sqrt{2\pi S(\beta m)}}
\]
represents the fraction of points in space that, when smoothed on a scale $\beta m$, have a overdensity $\ed_{\ell}$. The role of the second exponential in the fraction is to correct for points in space that are in fact part of a collapsed structure on a smoothing scale larger than $\beta m$. In random walk theory, this is referred to as the existence of an absorbing boundary at the collapse overdensity $\ed_{\rm 0,c}(a)$. For scales much larger than galaxy clusters themselves, it can be shown that the second exponential, already for $\beta >1.2$ (for reference, turnaround is predicted to be around $\beta >1.6$) is less than $20\%$ of the first term. Dropping it can significantly simplify the final expression, and we do so for the remainder of this paper.

{\bf Simplification 2: first differentiate, then convert to non-linear overdensities.}
As discussed earlier, Eq.~(\ref{dd}) depends on the linearly extrapolated overdensity. Strictly speaking, in order to derive the most probable profile, we must first perform a change of variables using Eq.~(\ref{magic}) to recast the double distribution in terms of $\delta_\ell$, and then differentiate to obtain the value of $\delta_\ell$ as a function of $\beta m$ where the distribution reaches its maximum. However that would significantly complicate both the algebra involved, and the final resulting expression. Instead, we will adopt the approximate path of first calculating a most-probable profile of linearly extrapolated over-densities by differentiating Eq.(\ref{dd}), and then use the conversion relation (eq. ~(\ref{magic})) to convert the profile of linearly extrapolated overdensity to profile of non-linear overdensity. 
Physically, this would be equivalent to determining what the most probable density profile around a structure that has collapsed today looked like in the early Universe, when the density field was still in the linear regime; and then, evolving that early most-probable profile forward in time, using the spherical collapse model. We expect some inaccuracy - which we will evaluate through comparison with simulations - to stem from this choice, since the  “most-probable profile” is a statistical quantity and does not represent, as a whole, the density profile around a single real structure (see discussion of the "typical profile" in \citealp{Betancort-RijoEtal06},  and also in  \citealp{Barkana} for a discussion of similar strategies). 

Differentiating Eq.~(\ref{dd}) with respect to $\ed_\ell$, using the simplifications discussed above, and setting the result equal to zero returns the following simple expression for the most probable profile of overdensity linearly extrapolated to the present cosmic time: 
\begin{equation}\label{attoday}
    \edl =\ed_{0,c}(a)\frac{S(\beta m)}{S(m)}\,.
\end{equation}
We note that $\ed_{0,c}(a)$ is the overdensity of a structure collapsing at time $a$, linerarly extrapolated to $a=1$ (today). It is related to $\ed_{c}$ (the overdensity of a structure collapsing at $a$ extrapolated to time $a$, a universal value for all epochs $a$) through $\ed_{0,c}(a) = \ed_{c}D(1)/D(a)$, where $D(a)$ is the linear growth factor at redshift $a$. Using this relation, we can rewrite Eq.~(\ref{attoday}) in terms of quantities 
extrapolated to the time of collapse of the central structure $a$:
\begin{equation} 
\tilde{\delta}_{\ell,a}  =\tilde{\delta, c}\frac{S(\beta m)}{S(m)}.
\end{equation}
Finally, we use Eq.~(\ref{magic}) to convert the 
$\edl{}_a$
to $\delta_\ell(a)$. This is now possible,  
because both the linearly extrapolated and the nonlinear (spherical collapse) overdensities are calculated at the same time $a$. This yields
\begin{equation}
    \ed_c \left[
    1-\left(1+\delta_\ell\right)^{-1/\ed_c}
    \right] = \ed_c
    \frac{S(\beta m)}{S(m)}\,.
\end{equation}
Solving for $\delta_\ell$ we get 
\begin{equation}
    1+\delta_\ell = \left[1-\frac{S(\beta m)}{S(m)}\right]^{-\ed_c}\,,
\end{equation}
which we can confirm has the correct asymptotic behavior, going to $\infty$ when $\beta =1$, and going to 1 (i.e., $\delta_\ell=0$, average matter-density of the universe) when $\beta \rightarrow \infty$.  We can also express this equation in terms of 
average densities within a sphere including mass $\beta m$ (instead of overdensities), 
\begin{equation}\label{enclosed}
    \rho_{\rm avg} (\beta)=  \rho_m \left[1-\frac{S(\beta m)}{S(m)}\right]^{-\ed_c}\,,
\end{equation}
where $\rho_m$ is the average matter-density of the Universe at the desired redshift.

We can simplify Eq.~(\ref{enclosed}) even further if the variance of the density field can be approximated as a power law in mass, $S(m) = m^{-\gamma}$, with \(-\gamma = d\ln S /d \ln m \) (in Appendix \ref{sigma} we explore the legitimacy of this assumption). With this substitution, we obtain
\begin{equation}\label{enclosed_beta}
    \rho_{\rm avg} =  \rho_m \left[1- \beta^{-\gamma}\right]^{-\ed_c}\, = \rho_{m,0}a^{-3} \left[1- \beta^{-\gamma}\right]^{-\ed_c}.
\end{equation}
Put in words, up to the validity of all the approximations we have discussed so far, the average matter density of a sphere enclosing mass $\beta$ times central galaxy cluster mass (so $\beta >1$ always), normalized to the mean matter-density of the universe at the time of observation, is universal, and equal to $(1-\beta^{-\gamma})^{-\ed_c}$. For very large $\beta$, this profile asymptotes to 1 (i.e., eventually the matter density of increasingly large spheres encompassing the cluster eventually tends to the average matter-density of the Universe at the time of observation). For $\beta \rightarrow 1$, the profile asymptotes to  $\infty$, as is the expectation in the spherical collapse model, where a structure collapses to a singularity. In practice, this means that the validity of this (outer-region) profile will break down at some $\beta >1$, to be determined by comparison with simulations.  


\subsection{Mass scalings} \label{Mass scalings}

The very existence of a universal average density profile dependent only on $\beta$ (and not on $m$) implies that the virial mass will scale with any other "constant overdensity" mass that is larger than the virial mass itself. If, then, we take the virial (collapsed, relaxed) mass of the central cluster to be the usually-assumed $m_{200}$ (i.e., a mass enclosed within a sphere which is, on average, 200 times denser than the average background-matter density of the Universe), and $m_{X}$ to be a different fixed-overdensity mass (a mass enclosed in a sphere $X$ times overdense with respect to the background Universe, with $X<200$), it is straightforward to derive a scaling between the two.  In order to see this we solve Eq.~(\ref{enclosed_beta}) for $\beta$ to get: 
\begin{equation}
    \beta = 
    \left[
    1 - \left(\frac{\rho_{\rm avg}}{\rho_{m} (a)}\right)^{-1/\ed_c}
    \right]^{-1/\gamma}\,.
\end{equation}
So now any constant overdensity criterion $\rho_X = X\rho_m(a)$ will yield a specific value for $\beta$ (a value independent of $m$, and so the same for clusters of all masses) :
\begin{equation} \label{eq. 10}
    \beta_X= 
    \left[
    1 - X^{-1/\ed_c}
    \right]^{-1/\gamma}\,.
\end{equation}
And thus the sphere of average overdensity $\rho_X = X\rho_m(a)$ will most probably contain a mass $\beta_X m$ ("most probably" because our formalism yields the most probable average density profile). Then, the most probable scaling of $m_X$ with $m$ (i.e., most commonly, with $M_{200}$),  will be 
\begin{equation} \label{scaling_relation}
 m_x = \beta_X m = 
    \left[
    1 - X^{-1/\ed_c}
    \right]^{-1/\gamma} m \,.
\end{equation}
This will also hold for the turnaround mass, since the turnaround density is a constant for a given cosmology and a given redshift (for example, for the concordance $\Lambda$CDM cosmology today, $X=11$ for the turnaround mass). 
 \begin{equation}
 M_{\rm ta}  =   \left[
    1 - X_{\rm ta}^{-1/\ed_c}
    \right]^{-1/\gamma} M_{200} \,.
\end{equation}
This means that we have a theoretically predictable scaling between $M_{200}$ and $M_{ta}$, for any cosmology. Because $\gamma$ has a slight dependence on mass, the logarithmic scaling between masses is expected to have some tilt compared to the the $M_{\rm ta} \propto M_{200}$ line.

Similarly, we can derive a scaling between any two overdensity masses, say $m_X$ and $m_Y$, defined as the masses of spheres with average matter density $X\rho_{m}(a)$ and $Y\rho_{m}(a)$, respectively, provided that both $X$ and $Y$ are lower than 200 (or whatever the threshold for the virialized part of the cluster is taken to be):
\begin{equation} \label{eq.13}
m_X = \left(\frac{1-X^{-1/\ed_c}}{1-Y^{-1/\ed_c}}
\right)^{-1/\gamma}m_Y\,.
\end{equation}
Detailed recipes for deriving the values of  $\gamma$ and $\ed_c$ used in this work are given in appendices \ref{sigma} and \ref{edc}, respectively.
\section{Simulations and methods} \label{section 3}
In this section, we discuss the simulations used for testing the validity and accuracy of the analytic profile and the predicted mass scalings. We present the criteria we adopted to select our halo sample, and the methodology we used to construct the most-probable average mass profiles for these halos. 

\subsection{Simulation overview}

To validate our analytical models against realistic, simulated average mass profiles of dark matter halos in cluster-sized systems, we  used the same cosmological simulations as in our prior work ( \cite{Korkidis_etal2023}; a more in depth presentation of the simulation can be found therein). This includes the large-box concordance $\Lambda$CDM MDPL2 simulation (1000 $h^{-1}$ comoving Mpc on a side, particle mass $1.51 \times 10^9 h^{-1} {\rm M_\odot}$), and the Virgo consortium simulations, with identical $239.5 h^{-1}$Mpc - sided boxes for three different cosmologies [ a concordance $\Lambda$CDM, a no-$\Lambda$ $\Omega_m=0.3$ (OCDM), and a "standard" $\Omega_m=1$ flat CDM (SCDM)], and particle masses equal to $6.86 \times 10^{10} h^{-1} {\rm M_\odot}$ for the $\Lambda$CDM/OCDM, and $22.7 \times 10^{10} h^{-1} {\rm M_\odot}$ for the SCDM respectively. 

\subsection{Building the profiles}
The profile that we derived in Section \ref{section 2}, described the most probable profile of the average matter density within concentric spheres of increasing enclosed mass, as a function of that enclosed mass. To compare the analytical predictions against the simulated data, we analyzed the matter distribution around 1000/900 randomly selected galaxy-cluster-sized (\(\rm M_{200} \geq 8\times 10^{13} \ M_{\odot}\)) dark matter structures from each of the MDPL2/Virgo simulations, respectively. Given that in hierarchical structure formation cosmologies the halo distribution follows a mass function in which larger structures are more rare, during our sampling we divided the halo catalog in 30 mass bins, and from each bin, we selected 40 random clusters. We segmented the region surrounding the center of each galaxy cluster into 500 concentric spheres extending up to \(\rm 10\times R_{200,m}\) in radius, ensuring that each halo was contained within the simulation's boundaries. Within each sphere, we computed the enclosed mass, and, dividing by the volume of the sphere, the corresponding dark matter density. The chosen number and size of bins were sufficiently large to accurately represent the profiles at all enclosed overdensity masses for both group and cluster-sized halos.
Finally, we normalized the enclosed mass for each individual halo profile to its virial mass, specifically $\rm M_{200}$.

\section{Results} \label{section 4}
\subsection{Mass density profile of simulated halos}
\begin{figure}[htb!]
    \centering
    \includegraphics[width=1\columnwidth]{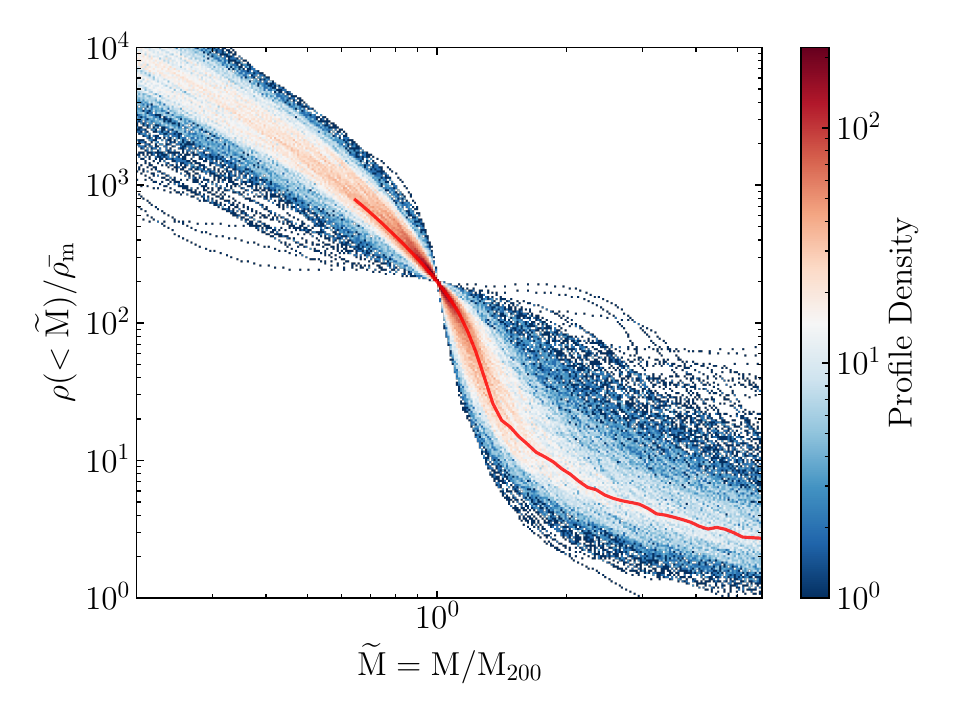}
    \caption{Average matter density (in units of the background-Universe average matter density) within concentric spheres around MDPL2 $z=0$ cluster-sized halos, as a function of mass that is enclosed in each sphere. The colorscale encodes the number of lines per pixel on the plot. The enclosed mass is normalized to \(\rm M_{200}\), so all profiles converge at an average density 200 per the definition of \(\rm M_{200}\). The red solid line depicts the profile mode. The independent variable $\tilde{M}=M/M_{200}$ is equivalent to  $\beta$ in the analytic profile of Eq.~\ref{enclosed_beta}.}
    \label{Fig. bundle}
\end{figure}

We first confirmed that the type of profiles examined here do cluster around a most probable behavior which our analytical profile aims to model. 
Figure \ref{Fig. bundle} presents the compiled profiles from the MDPL2 simulation at $z=0$. The colorbar indicates the density of lines on the plot. This visualization clearly shows that most profiles tend to converge around a common profile. In Fig.~\ref{Fig. bundle} the independent variable $\rm \tilde{M}=M/M_{200}$ is equivalent to  $\beta$ in the analytic profile of Eq.~\ref{enclosed_beta}, assuming that $\rm M_{200}$ is a reasonable approximation of the true relaxed mass $m$ of the central cluster.

To formally capture and represent this pattern, we grouped these normalized profiles into 40 linear mass bins, from 0 to \(\rm 4\times M_{200}\). For each bin, we determined the most probable value (mode) of the density. This was achieved by employing a Gaussian kernel density estimator \footnote{For the bandwidth determination the algorithm that we employed used Scott's rule \citep{Scotts_rule}.} \citep{SciPy}  to model the probability density function (PDF) of densities in the bin, and subsequently identifying the density for which the PDF is maximum. The resulting mode profile is depicted by the red solid line in Fig. ~\ref{Fig. 1.}, closely tracking regions with high profile density, as anticipated.

\begin{figure}[htb!]
    \centering
    \includegraphics[width=1\columnwidth]{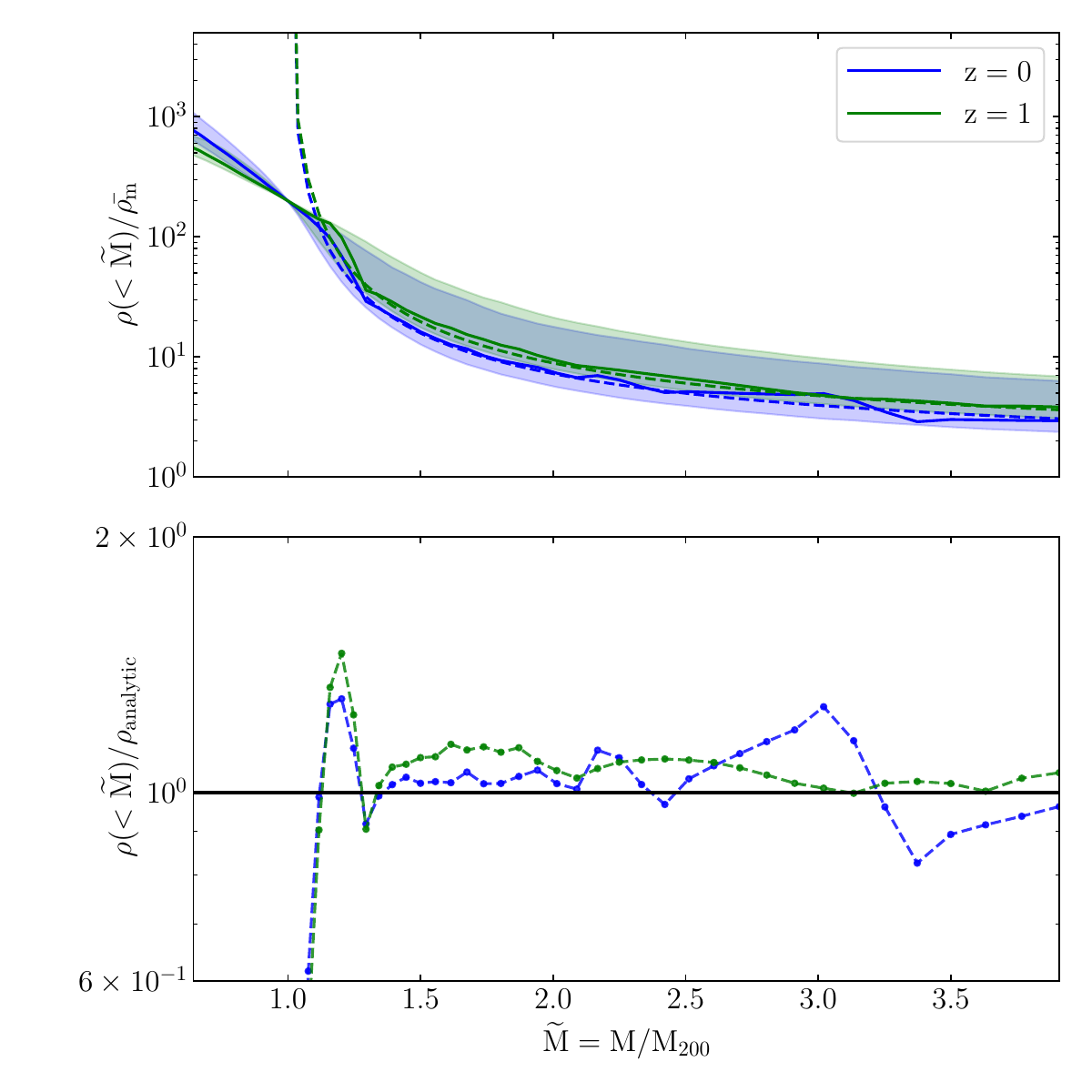}
    \caption{Performance of our analytic density profile for different redshifts. Upper panel: most probable (mode) average matter density within spheres of enclosed mass $M$ as a function of $M$ (normalized to $M_{200}$; $\tilde{M}$ is equivalent to $\beta$ in Eq.~(\ref{enclosed_beta})), from our analytic result (Eq.~\ref{enclosed_beta}, dashed lines), and from the MDPL2 simulation halos (solid lines; the shaded region includes 68\% of the densities PDF), for $z=0$ (blue) and $z=1$ (green). Lower panel: ratio of MDPL2 over analytic mode profile as a function of enclosed mass. In the bottom panel, the ratio of the mode to the analytic profiles is displayed. 
    }
    \label{Fig. 2.}
\end{figure}

We next tested the extent to which our analytic profile matches the most probable profile seen in simulations.
In Fig.~\ref{Fig. 2.}, we overlay the analytical profile with the mode of the profiles derived from the MDPL2 simulation. Different curve colors stand for different redshifts of our halo samples. For the analytical profile, we incorporate the cosmological parameters from the simulation ($\rm \Omega_m, \Omega_{\Lambda}, h_0, n_s$) and the median value of $\rm M_{200}$ for the halo sample (\(\rm 6.3\times 10^{14}\ M_{_\odot}\) for \(\rm z=0\) halos and \(\rm 1.2\times 10^{14}\ M_{_\odot}\) for \(\rm z=1\)). The lower panel compares the simulated and analytical profiles by examining their ratio.

The analytical profiles, developed using the spherical collapse model to evolve linearly extrapolated overdensities into the non-linear regime, tend to infinity as the enclosed mass approaches $\rm M_{200}$ \footnote{Under our assumptions, the model assumes virialization for simulated clusters when the density is 200 times the background mass density, a consequence of employing spherical top hat thresholds; for a recent discussion see \cite{Delos}.}. For $\tilde{M} \equiv \beta \geq 1$, the analytical profile shows an excellent alignment with the mode profile; fluctuations are within 15\%. The profile exhibits little variation with redshift, beyond what is encoded in the overall increase of average background matter density with increasing redshift.

\begin{figure}[htb!]
    \centering
    \includegraphics[width=1\columnwidth]{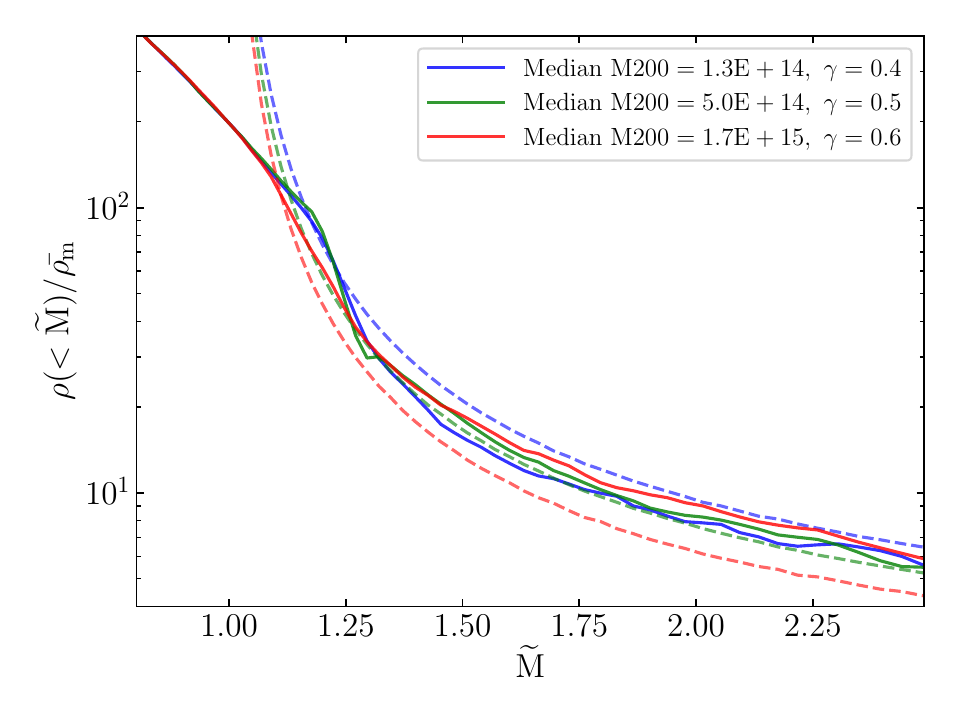}
    \caption{Most probable average matter-density profile from our analytical model (dashed lines) and MDPl2 (solid lines), for different ranges of cluster masses \(M_{200}\). The independent variable $\tilde{M}$ is equivalent to $\beta$ in Eq.~(\ref{enclosed_beta}).
    }
    \label{Fig. 3.}
\end{figure}
Simulations consistently show that the density profile around clusters exhibits near-universal characteristics when normalized against a constant overdensity radius or mass relative to the background. This behavior is replicated in Figure \ref{Fig. 3.}, where, with solid lines, we plot the mode density profiles of MDPL2 clusters at \(z=0\), against their enclosed mass in concentric spheres of increasing radius. Different colors correspond to different average masses of the halo sample. The consistency of the profiles between different mass bins is even stronger in simulations than in our analytical profiles (shown with dashed lines), with deviations between simulations and analytic profile for the the smaller and larger mass bins reaching 25\%. This variation of the analytic profile with cluster mass originates in the mild dependence of $\gamma$ on mass (each profile in the figure has a different $\gamma$ value as shown in the legend; also, see appendix \ref{sigma} and Fig.~\ref{Fig. 1.}). 
We have confirmed that the expected mild mass-dependence of the analytic mode profile is real, and not a result of our approximations: it persists when the full expression for $\rm S(m)$ is used rather than the power-law approximation: the deviation between the two profiles is negligible for the range of $\rm \tilde{M}$ considered here, when $\gamma$ is taken to be equal to $-d{\rm ln} S/d {\rm ln} m |_{\rm m=M_{200}}$ (as we do in all cases, using the median $\rm M_{200}$ of our sample each time). For much higher $\rm \tilde{M}$, of course, when the transfer function $\rm T(k) \rightarrow 1$, $\rm S(m)$ does asymptote to a power-law (see Eq.~\ref{TheSofMEq}), and the full universality of the mode profile is recovered; however this becomes relevant for mass scales much larger than those considered in this paper.  The dependence of the mode profile also persists if we replace the approximation of Eq.~(\ref{magic}) with the more accurate but also more cumbersome Eq.~(8) of \cite{ShethTormen02}, also used by \cite{PradaEtal06}.


The lack of such dependence in simulations is plausibly an effect of the somewhat fuzzy correspondence between $\rm M_{200}$ and the true "collapsed mass" entering the analytical mode profile. Even under the assumption of spherical symmetry, the SCM is only applicable for shells that have never undergone shell crossing. In this context, the appropriate collapsed mass is somewhat ambiguous. Here we have taken it to be $\rm M_{200}$. However, if one were to require that all mass that has undergone shell crossing is considered to be part of the collapsed mass, then the splashback mass \citep{Splashback1, Splashback5, Splashback6} should also be considered. This is likely one of the effects that partly contribute to the offset observed in Fig. ~\ref{Fig. 3.} between the solid and the dashed blue curves.

This effect would also be exaggerated by the underlying assumption in Eq.~\ref{dd} that steps in enclosed mass result in uncorrelated corresponding steps in enclosed underdensity (implemented formally by the adoption of a sharp-in-k filter for S(m); see also discussion in Appendix A). This assumption is required for the excursion-set formalism used to derive Eq.~\ref{dd} to be strictly applicable; however, in the analysis of simulations, successive steps in overdensity corresponding to successive steps enclosed mass are correlated (they are calculated using a top-hat, rather than a sharp-in-k, filter). The extent of this mismatch would be dependent on the mass accretion rate.

Finally, another factor to consider is the finite accuracy with which we determine the mode density in every sphere of a given $\rm \tilde{M}$: the mode profile remains reasonably within uncertainties of the simulated mode across mass bins, simulations, and redshifts.

\subsection{Mass scalings} \label{mass_correlation}

\begin{figure}[htb!]
    \centering
    \includegraphics[width=1\columnwidth]{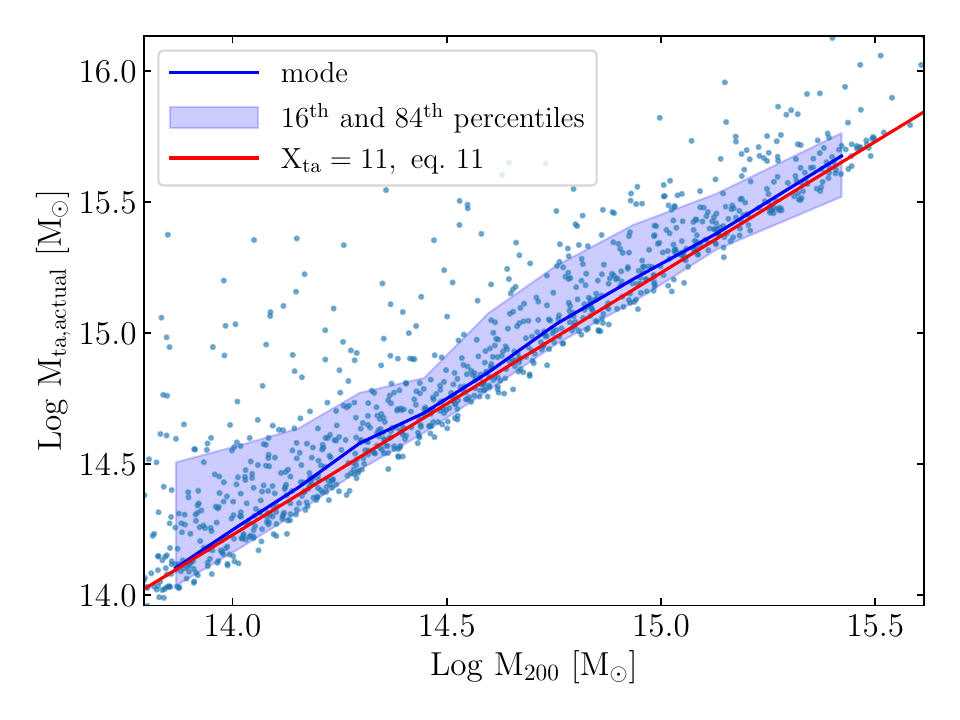}
    \caption{Correlation between turnaround mass \(\rm M_{ta,actual}\) and collapsed mass \(\rm M_{200}\) for the MDPL2 cluster sample at \(\rm z=0\). The red solid line depicts the theoretical scaling relation of Eq.\ref{scaling_relation}. The blue solid line represents the mode value of the blue points in bins of \(\rm M_{200}\). The blue shaded region corresponds to the 16th and 84th percentiles}
    \label{Fig. scaling.}
\end{figure}

\begin{figure*}[htb!]
    \centering
    \includegraphics[width=0.48\textwidth]{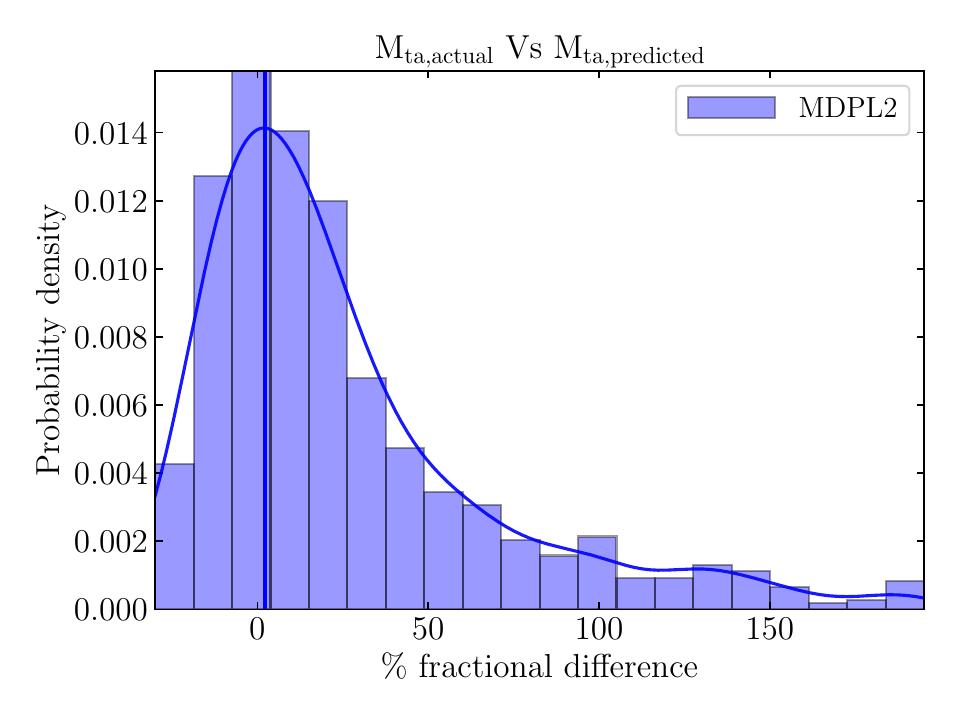}
    \hfill
    \includegraphics[width=0.48\textwidth]{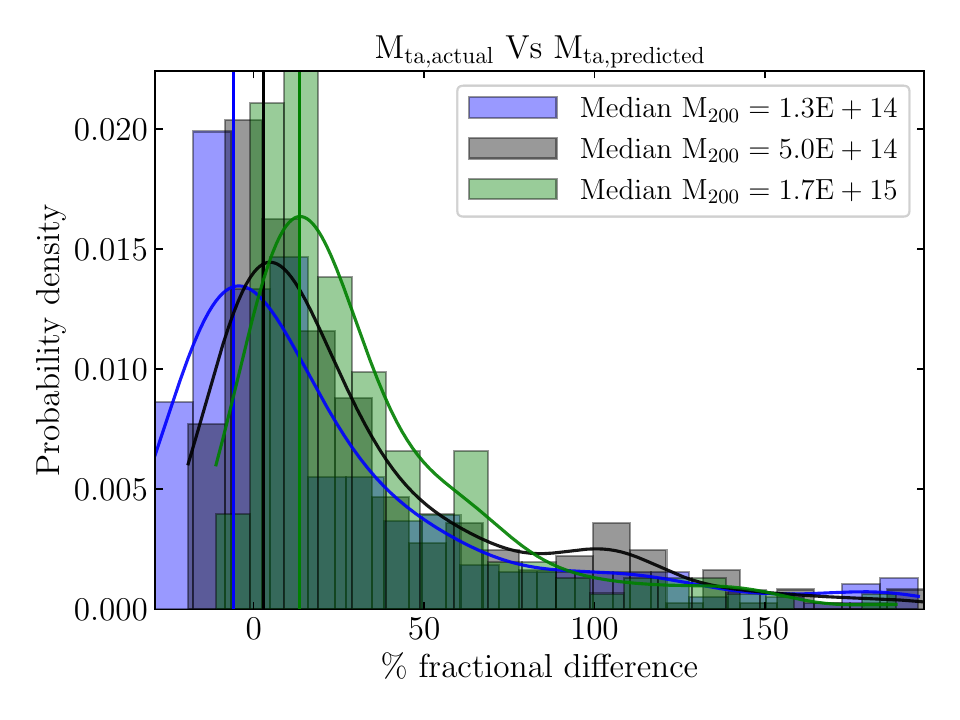}
    \caption{Probability density of the fractional (percentage) difference between the turnaround mass (here $M_{11}$) predicted from the scaling relation of Eq.~(\ref{scaling_relation}) and its actual value, for the full  MDPL2 simulated cluster sample at \(\rm z=0\) (left panel), and for halo subsamples 
    of different median $M_{200}$ (right panel). 
    In each case, the solid line shows a Gaussian kernel estimator of the probability density function, and the
    vertical line marks the mode of the error distribution. 
The scaling relation tends to overpredict $M_{11}$ for lower $M_{200}$ and underpredict it 
for higher $M_{200}$. 
}
    \label{Fig. 4.}
\end{figure*}

We now turn to evaluate the performance of our scaling relation  between different overdensity masses (Eq.~ \ref{scaling_relation}) for our halo samples. As detailed in our prior works \citep{KorkidisEtAl, Korkidis_etal2023}, the kinematically-defined turnaround mass $\rm M_{ta, kinematic}$ (the mass enclosed within the kinematically-identified turnaround radius) corresponds well to an overdensity mass, consistent with the prediction of the spherical collapse model. For concordance $\Lambda$CDM and $z=0$, spherical collapse predicts that $X_{\rm ta}=11$. We also set the 
collapsed mass $m=\rm M_{200}$. For each halo, employing Eq.(\ref{scaling_relation}), we can predict the turnaround mass $\rm M_{ta, predicted} \equiv M_{11, predicted}$ from its $\rm M_{200}$ and also calculate its actual mass within that overdensity ($X=11$, $\rm M_{11, actual} \approx M_{ta, actual}$).

\begin{figure*}[htb!]
    \centering
    \includegraphics[width=0.9\textwidth]{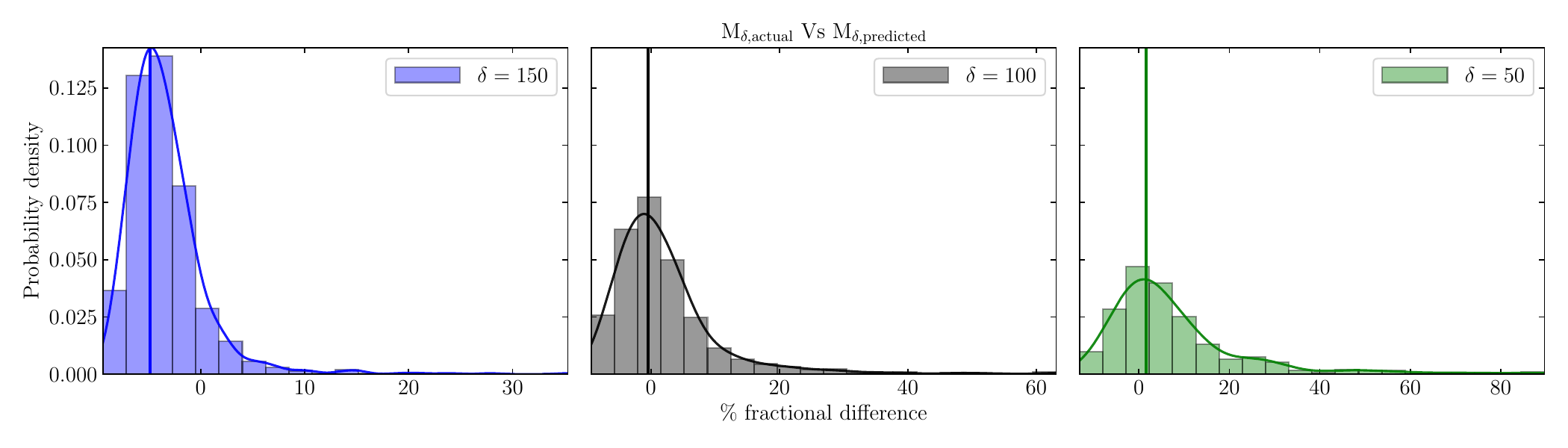}
    \caption{
    Performance of our scaling relation 
    for masses corresponding to different overdensity values $\delta$  (equivalent to $X$ in Eq.~\ref{scaling_relation}) in the MDPL2 sample. The solid curves and vertical lines are the same as in Fig.~
\ref{Fig. 4.}.
    As $\delta$ decreases (left to right), the most probable offset decreases, while high-error tails become more pronounced. Both behaviors can be traced to the performance of the analytic profile (see text for discussion). 
    }
    \label{Fig. 5.}
\end{figure*}

\begin{figure*}[htb!]
    \centering
    \includegraphics[width=0.9\textwidth]{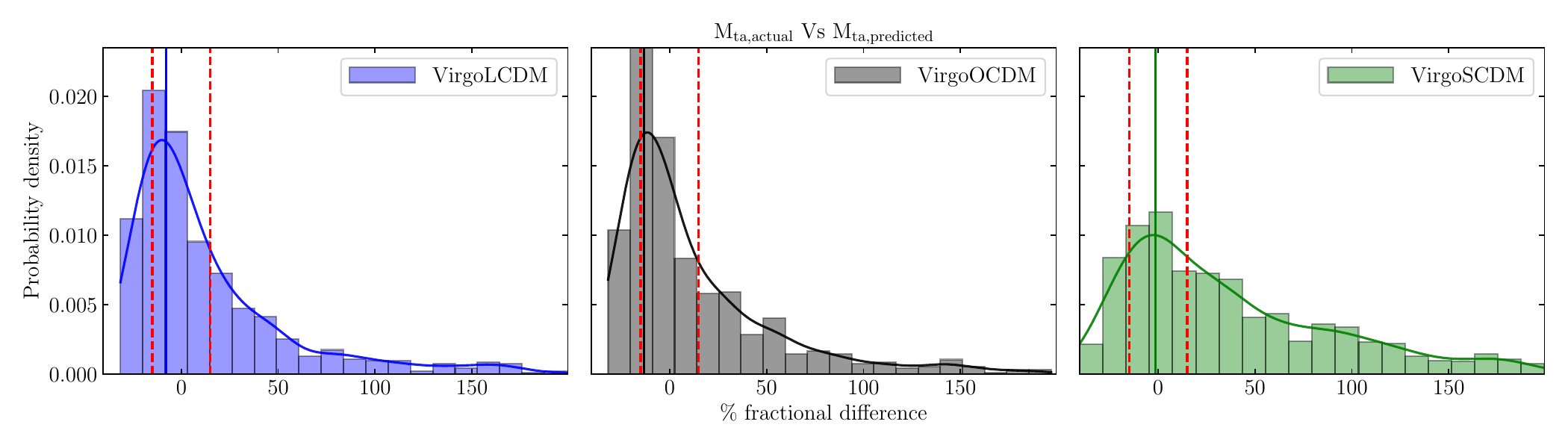}
    \caption{Performance of our scaling relation for three different Virgo simulations cosmologies. All panels correspond to \(\rm z=0\) snapshots. 
    The median \(\rm M_{200}\) of the halo samples are \(\rm 1.2\times 10^{14} M_{\odot}\) for VirgoLCDM/VirgoOCDM and \(\rm 1.7\times 10^{14} M_{\odot}\) for VirgoSCDM. Dashed red lines correspond to \(\pm 15 \%\) error.}
    \label{Fig. 6.}
\end{figure*}

We first test that such a scaling relation does exist in simulations, for the case that is of most interest to us (scaling of $\rm M_{200}$ with the turnaround mass). In Fig.~\ref{Fig. scaling.} we plot the turnaround mass of each cluster in our MDPL2 sample, \(\rm M_{ta,actual}\), as a function of its $\rm M_{200}$. The mode along with the 16th and 84th percentiles of this scaling in bins of $\rm M_{200}$ is shown with the solid blue line and the shaded region, respectively, and they do confirm that a tight scaling between the two masses is indeed present in the MDPL2 sample, although outliers do exist. For comparison, we overplot, with the red line, the mode scaling predicted by Eq.~(\ref{scaling_relation}), and it is already obvious that the analytic mode scaling is in excellent agreement with the mode scaling seen in simulations.
 
To further quantify the performance of our scaling we plot, with the histogram in the left panel of Figure \ref{Fig. 4.}, the fractional difference, as a percentage, between the turnaround mass $\rm M_{11, actual}$ measured from the simulation data, and the turnaround mass $\rm M_{11, predicted}$ predicted by our scaling relation using $M_{200}$ as input.  
The histogram has a pronounced peak near zero, suggesting that the scaling relation predicts the turnaround mass well. However, the distribution of residuals does feature a long tail, so the mean of this distribution has a non-negligible bias (up to 20\%). 


Another factor affecting the performance of the scaling relation is the mass range of the sample. This is explored in the right panel of Figure \ref{Fig. 4.}. Here, different colors correspond to subsamples of different mass ranges, with the median mass in each subsample shown in the legend. Although the qualitative features of the histograms remain consistent, we observe a systematic trend from negative to positive offsets of the peak of the distribution as the median sample mass increases. The absolute value of this offset is up to 15\% for the range of masses considered here. This behavior is anticipated based on the performance of the analytic profile for different sample mass ranges seen in Fig.~\ref{Fig. 3.}.

In Fig.~\ref{Fig. 5.} we evaluate again the performance of our mass scaling for different values of overdensity $\delta$ at which the mass is evaluated (corresponding to $X$ in Eq.~\ref{scaling_relation}). 
Each panel corresponds to a different value of $\delta$, which decreases from left to right, and which is displayed in the legend of each panel.  In all scenarios, the distribution peaks at $<10\%$ error. The most probable error increases with increasing overdensity, a behavior we anticipate due to the analytic profile diverging at the assumed collapse overdensity $\delta = 200$. However, the high-error outlier tails decrease with increasing overdensity, a behavior also anticipated from Fig.~\ref{Fig. bundle}: closer to $M_{200}$ individual profiles are more clustered around the mode profile. For this reason, deviations of individual outliers from the analytic mode profile, and the associated deviations from the mass scaling derived from the mode profile, are more pronounced for higher mass scales (lower overdensity values).  

The same analysis for $X \equiv X_{ta}$
 (turnaround mass) is then conducted in Figure \ref{Fig. 6.} for halo samples from three Virgo simulations, each corresponding to a different set of cosmological parameters: $\Omega_m=0.3, \Omega_\Lambda = 0.7$ (VirgoLCDM);  $\Omega_m=0.3, \Omega_\Lambda = 0$ (VirgoOCDM); and $\Omega_m=1, \Omega_\Lambda = 0$ (VirgoSCDM). In all cases, we analyzed \(\rm z=0\) snapshots.  The behavior in the first two cosmologies, featuring the same dark matter content at the present cosmic epoch, is very similar. This is consistent with our findings both analytically in the context of the spherical collapse model \citep{PavlidouEtal2020} and in simulations \citep{Korkidis_etal2023} that the behavior on turnaround scales at some specific redshift depends primarily on the average matter density at that same redshift.
The mild difference of the most-probable error compared to the MDPL2 sample (Fig.~\ref{Fig. 4.}) likely stems from the smaller box size in the Virgo simulations, which results to a smaller median mass of the halo sample. As also seen in Figure \ref{Fig. 4.}, in this range of halo masses the scaling tends to overpredict the turnaround mass (negative mode error).  
%
In the VirgoSCDM box, where the median mass of the sample is larger, the 
error mode is almost zero, and a more pronounced high-error tail is present, also consistent with our findings for different cluster mass ranges in MDPL2. It would appear that sample mass drives the performance of the scaling relation of Eq.~(\ref{scaling_relation}), more than the variation in cosmological parameters, which appear to be accounted for sufficiently by the model. 

\begin{figure}[htb!]
    \centering
    \includegraphics[width=1\columnwidth]{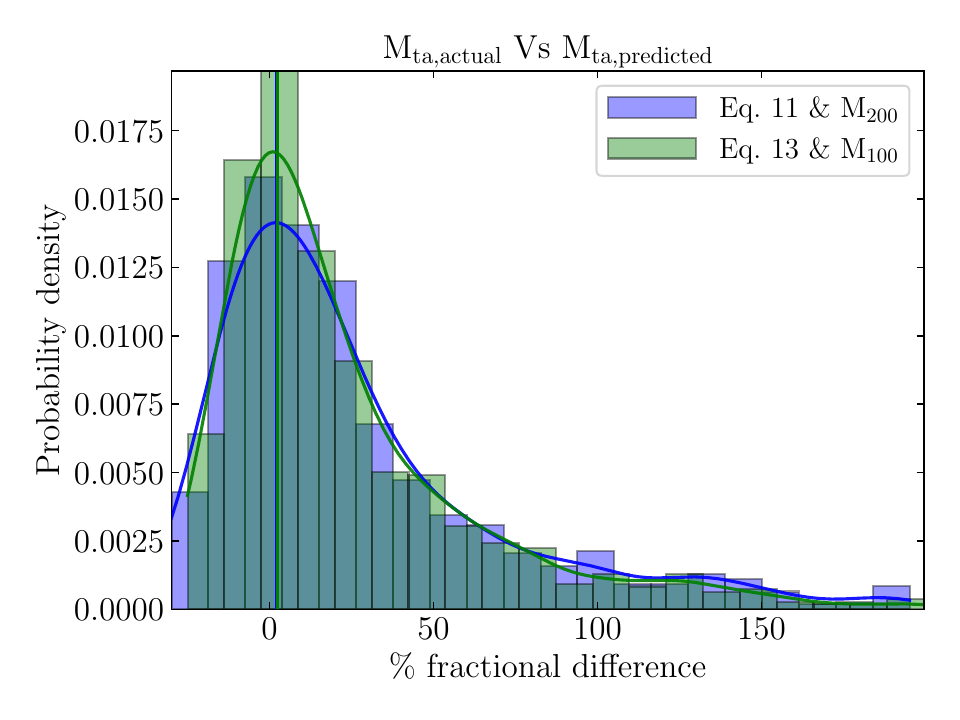}
    \caption{Comparison between the scaling relations delineated in Eq.~\ref{eq.13} and Eq.~\ref{scaling_relation}. The blue histogram depicted herein corresponds to that of the left panel in Fig.~\ref{Fig. 4.}. Superimposed on this histogram is the fractional difference between the actual value of \(\rm M_{ta}\) (\(\rm X=11\)) and its predicted counterpart assuming a prior knowledge of \(\rm M_{100}\) (that is, if we set \(\rm Y=100\)).}
    \label{Fig. C1.}
\end{figure}

Finally, we also test the performance of Eq.~ \ref{eq.13} - a scaling relationship between any two overdensity masses \(\rm m_X \ and \ m_Y \). We employ this scaling relation to estimate the actual value of the turnaround mass \( \rm M_{ta, actual} \ (X=11) \) using an overdensity mass \(\rm M_{100} \ (Y=100) \), and plot the distribution of its percentage difference from $M_{\rm ta, actual}$ in Fig.~\ref{Fig. C1.}. In the same figure, we also overplot the same distribution for the prediction derived from Eq.~(\ref{scaling_relation}).  The comparison indicates an almost identical position for the peak of the distributions; however, when the turnaround mass is derived from $M_{100}$ rather than $M_{200}$ the error distribution is tighter around zero - a result of the analytic profile performing better away from its divergence point at the collapse mass.

\section{Conclusions}\label{section 5}

In this study, we derived a theoretical model for the most-probable outer average density profile of large galaxy clusters as a function of enclosed mass, using excursion set theory. This model, based on Gaussian early universe statistics and the spherical collapse framework, has two parameters, $\gamma$ and $\ed_c$, both of which can be analytically derived from the cosmological parameters ($\Omega_m$, $\Omega_\Lambda$, and power spectrum slope $n$), and the median virial mass of the sample under consideration. 
We compared our profile with profiles around galaxy-cluster--sized halos in
\(\rm \Lambda CDM\) simulations, 
and found good agreement across mass ranges and redshifts. 

From the profile we derived a scaling relation that links different overdensity masses, and found that it performs well with the the peak of the error distribution remaining below $15\%$ for all cosmologies, sample masses, and overdensities we tested. 
We have traced residual inaccuracies of the scaling relation to two factors. 

The offset from zero of the maximum of the error is dependent on the exact cluster mass range considered. In particular, the mode profile in simulations appears to be more universal (similar across mass ranges) than the analytic model predicts. This is plausibly an effect of an imperfect identification between the collapsed mass $m$ of the analytic profile and the overdensity-200 mass $M_{200}$ in simulated clusters, leading to a cross-contamination of mass ranges. 

The tail of the error distribution tends to grow as the mass scale we consider grows (or, equivalently, as the overdensity where the mass is calculated decreases). This is due the increasing spread of the profile distribution with increasing scale. As a result, masses at overdensities closer to 200 (say, 150 or 100) scale more tightly with $M_{200}$ than the turnaround mass ($M_{11}$ for $z=0$ in concordance \(\rm \Lambda\) CDM). 

Our analytical model demonstrates the existence, in principle, of a scaling between any two overdensity masses, dependent on cosmology and redshift, and it provides a reasonable approximation to this scaling directly from these parameters. A fit of the parameter $\gamma$ to a particular simulated sample of clusters with masses in the range of interest can provide an even more accurate scaling should one be required. We have confirmed that such fits return values of $\gamma$ that always fall within those predicted theoretically (see appendix A and Fig.~\ref{Fig. 1.}) for masses in the galaxy-cluster range. 



\begin{acknowledgements}
We would like to thank the anonymous referee for his constructive report. We acknowledge support by the Hellenic Foundation for Research and Innovation under the “First Call for H.F.R.I. Research Projects to support Faculty members and Researchers and the procurement of high-cost research equipment grant”, Project 1552 CIRCE (GK, VP); and by the Foundation of Research and Technology – Hellas Synergy Grants Program (project MagMASim, VP). 
The research leading to these
results has received funding from the European Union’s Horizon 2020 research and innovation programme under the Marie Skłodowska-Curie RISE action, Grant
Agreement n. 873089 (ASTROSTAT-II).
The CosmoSim database used in this paper is a service by the Leibniz-Institute for Astrophysics Potsdam (AIP). The MultiDark database was developed in cooperation with the Spanish MultiDark Consolider Project CSD2009-00064. The authors gratefully acknowledge the Gauss Centre for Supercomputing e.V. (www.gauss-centre.eu) and the Partnership for Advanced Supercomputing in Europe (PRACE, www.prace-ri.eu) for funding the MultiDark simulation project by providing computing time on the GCS Supercomputer SuperMUC at Leibniz Supercomputing Centre (LRZ, www.lrz.de). The Bolshoi simulations have been performed within the Bolshoi project of the University of California High-Performance AstroComputing Center (UC-HiPACC) and were run at the NASA Ames Research Center. The Virgo Consortium simulations used in this paper were carried out by the Virgo Supercomputing Consortium using computers based at Computing Centre of the Max-Planck Society in Garching and at the Edinburgh Parallel Computing Centre. The data are publicly available at www.mpa-garching.mpg.de/galform/virgo/int\_sims. Throughout this work we relied extensively on the PYTHON packages Numpy \citep{Numpy}, Scipy \citep{SciPy} and Matplotlib \citep{Matplotlib}.

\end{acknowledgements}

%
%

\bibliographystyle{aa}
\bibliography{bibliography}

\begin{appendix} 

\section{Density field variance} \label{sigma}

In deriving Eq.~(\ref{enclosed_beta}) from Eq.~(\ref{enclosed}) we assumed that, for a limited range of masses, $S(m)$ can be approximated by a power law, that is

\begin{equation}\label{forS}
    \frac{S(\beta m)}{S(m)} \approx \beta^{-\gamma}
\end{equation}
 where 
 \begin{equation}
-\gamma = \frac{d\ln S}{d \ln m}.
\end{equation}

Here, we evaluate the validity of this assumption, and we provide a recipe for the evaluation of $\gamma$. The variance of the density field is given by 
\begin{equation}\label{TheSofMEq}
S(m) = \int_{k=0}^{k(m)}\!\!\!k^2dk |\delta_k|^2 = \sigma_8^2 
\frac{\int_{k=0}^{k(m)} T^2(k)k^{n+2}dk} 
{\int_{k=0}^{k(m_8)} T^2(k)k^{n+2}dk} \,
\end{equation}
In this equation, we have made two assumptions. First, that right after matter-radiation equality, $\langle|\delta_k|^2\rangle$ can be simply described in terms of a power-law in $k$ modified by a transfer function,
$\langle|\delta_k|^2\rangle \propto T^2(k)k^n$. 
Second, that the window function for the calculation of the variance is sharp in $k-$space. This assumption is necessary for random-walk formalism (used to derive Eq.~\ref{dd}) to be strictly applicable: each "step" in the time-like variable (here $S(m)$) should produce a step in the space-like variable (here, the overdensity corresponding to the mass scale $m$) that is completely independent from  the previous step. This requires that the k-modes producing a change in the spacelike variable not appear in any of the previous steps. A sharp-in-$k$ window function enforces this condition.  For a more extensive discussion on this assumption and its consequences see \citet{Bond91,LaceyCole93, PF05}.

The variance of the field has been normalized to the present time at scale of 8 comoving Mpc ($\sigma_8$).
For $T(k)$ we adopt the fitting formulae of \citet{Bardeen} for the adiabatic cold dark matter transfer function (Eq.~G3). We have verified that using other approximations (e.g., per \citealp{EisensteinHu}) does not alter substantially any of the results presented in this work. 

\begin{figure}[htb!]
    \centering
    \includegraphics[width=1\columnwidth]{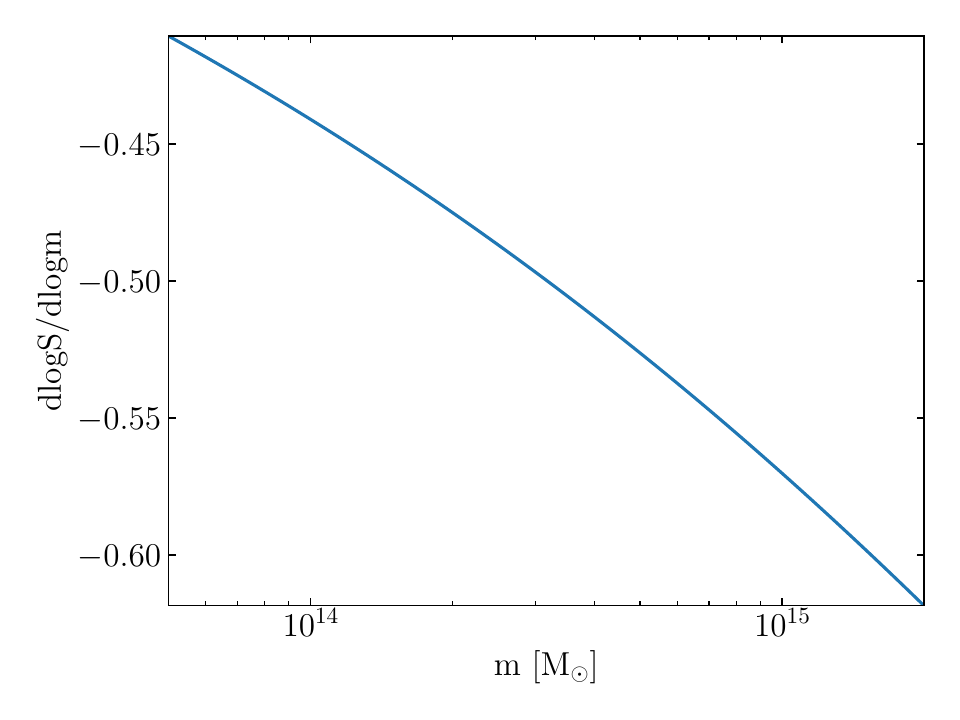}
    \caption{Variation in the logarithmic slope of the density field dispersion \(S\) for diverse values of the collapsed mass \(m\). The function's minimal curvature indicates near-consistency with a straight line across each order of magnitude in mass. This pattern suggests that approximating the variance of the density profile with a power-law model could be a plausible approach.}
    \label{Fig. 1.}
\end{figure}

To obtain $k(m)$, we integrate the sharp-in-k window function over all space and multiplying by $\rho_{\rm m,0}$ we obtain
\begin{equation}
k_c(m) = \left(\frac{6\pi^2\rho_{\rm m,0}}{m}\right)^{1/3}\,.
\end{equation}
Then, from Eq.~(\ref{TheSofMEq}), we can calculate $\rm S(m)$ and its logarithmic slope, which we plot in Fig.~\ref{Fig. 1.}. Clearly, this slope is not constant over the range of masses of interest (as it should be for a power law), but it does not vary wildly either. As a result, for a small range of masses, $S(m)$ can be approximated reasonably well by a power law. Of course as $m\rightarrow \infty$ the transfer function $T\rightarrow 1$ and $S(m)$ asymptotes to $\propto m^{-(n+3)/3}$, but this occurs on scales much larger than those of interest in this work. 

\section{Calculating $\ed_c$}\label{edc}
For a given cosmology, 
that is, a set of $\Omega_m$, $\Omega_\Lambda$, we first calculate the parameters
\begin{equation}
    \omega = \Omega_\Lambda/\Omega_m
\end{equation} and 
\begin{equation}
    \phi = (\Omega_m+\Omega_\Lambda-1)/\Omega_m\,.
\end{equation}
Note that for a flat cosmology, $\phi=0$, while for a cosmology without $\Lambda$,  $\omega=0$. 

Now let $a_p$ be the (evolving) radius of an initially overdense region (density perturbation) normalized so that, had the specific region begun its evolution at the same density as the bacground Universe, $a_p$ at the present cosmic epoch would have been equal to 1. The size of this perturbation at turnaround, $a_{pta}$, that is reaching collapse today (i.e. at scale factor of the background Universe equal to 1; note that this structure must have reached turnaround at sometime in the past), can be found through (e.g., \citealp{PF05, PavlidouEtal2020})
\begin{equation*}
    \int_{0}^{1} \frac{\sqrt{y}dy}{\sqrt{\omega y^3 - \phi y +1}} = 2a_{pta}^{3/2}\int_0^1\frac{\sqrt{z}dz}{\sqrt{\omega a_{pta}^3 z^3 - (\omega a_{pta}^3+1)z +1}}\,.
\end{equation*}
Then, the extrapolated overdensity $\ed_c$ to the time of collapse\footnote{In the case considered, which is of a structure achieving collapse today, that time is the time of scale factor equal to 1.} will be \citep{PF05}:
\begin{equation}
    \ed_c = \frac{3\Omega_m[(\omega a_{pta}^3+1)/a_{pta} - \phi]}{2}D(a_0), 
\end{equation}
where $D(a_0)$ is the linear growth factor 
for the present cosmic epoch $a_0=1$  \citep{Carroll1992}
\begin{equation}
D(a_0)= \Omega_m^{-1/2} \sqrt{1+\omega-\phi}
\int_0^{1}\left[
\frac{x}{1+\omega x^3 -\phi x}
\right]^{3/2}dx\,.
\end{equation}
Overall, the variations in $\ed_c$ in different cosmologies are small. For concordance $\Lambda$CDM ($\Omega_m = 0.3$, $\Omega_\Lambda=0.7$), $\ed_c=1.6757$. For SCDM ($\Omega_m=1$, $\Omega_\Lambda=0$), $\ed_c=1.6865$.

\end{appendix}

\end{document}